\documentclass[11pt]{article}
\usepackage{amsfonts}
\usepackage{mathrsfs}
\usepackage{stmaryrd}
\usepackage{amsmath}
\usepackage{amsthm}
\usepackage{amstext}
\usepackage{amsopn}
\usepackage{color}
\usepackage{graphicx}
\usepackage{epstopdf}
\usepackage{multirow}
\usepackage{enumerate}

\numberwithin{equation}{section}

\theoremstyle{plain}

\title{A DFA-based bivariate regression model for estimating the dependence of PM2.5 among neighbouring cities}
\author{Fang Wang$^{1,2}$, Lin Wang $^2$, and Yuming Chen$^3$\\
$^1$ College of Science,
 Hunan Agricultural University,
  Changsha, P. R. China 410128\\
  $^2$ University of New Brunswick, Fredericton, NB,
E3B 5A3, Canada\\
$^3$ Wilfrid Laurier University, Waterloo, ON, Canada N2L 3C5}
\date{}

\begin{document}

\maketitle

\begin{abstract}
On the basis of detrended fluctuation analysis (DFA), we propose a new bivariate linear regression model. This new model provides estimators of multi-scale  regression coefficients to measure the dependence between variables and corresponding variables of interest with multi-scales. Numerical tests are performed to illustrate that the proposed DFA-bsaed  regression estimators are capable of  accurately depicting the dependence between the variables of interest and can be used to identify different dependence at different time scales. We apply this model to analyze the PM2.5 series of three adjacent cities (Beijing, Tianjin, and Baoding) in Northern China. The estimated regression coefficients confirmed the dependence of PM2.5 among the three cities and illustrated that each city has different influence on the others  at different seasons and at different time scales. Two statistics based on the  scale-dependent $t$-statistic and the partial detrended cross-correlation coefficient are used to demonstrate  the significance of the dependence. Three new scale-dependent evaluation indices show that the new DFA-based bivariate regression model can provide rich  information on  studied variables.


\end{abstract}

\section{Introduction}

In recent years, air pollution has become a  more and more serious problem around the world. The new air quality model presented by the World Health Organization in 2016 confirmed that 92\% of the world's population lives in areas where air quality levels exceed their limits~\cite{NR}. Fortunately, more and more governments have realized the importance of managing air pollution and some actions have been placed. Nowadays, a common topic around the world is the  governance of the air pollution source  such as smog (the main ingredient is fine particulate matter). Many researchers have been involved in the study on the cause and propagation of smog~\cite{WH,HZ1,HZ2,HZ3,SL,SK,SC,ZZ}. Modern statistic methods provide some new perspectives to assess smog trends and propagation characteristics. Among them, most studies have focused on studying the correlations among various air pollution indicators  including  air pollution index (API), air quality index (AQI), fine particulate matter of PM2.5 (diameter $\leq 2.5\mu m$) concentrations, and PM10 (diameter $\leq 10\mu m$) concentrations, and very limited studies considered the correlations among neighboring areas. A common sense is that smog produced at one source place can spread to surrounding areas~\cite{SL,SK,SC,ZZ, WF0}. Therefore, it is more practical to explore the dependence of   air pollution indicators among adjacent cities as it helps assess the causes of local smog and its spread behavior. It has been found by a newly proposed time-lagged cross-correlation coefficient in Ref. \cite{WF0} that there are different degrees of correlation for PM2.5 series between four neighboring cities in Northern China. However, what has not been investigated is how the PM2.5 series of one city depends on those of the neighbouring cities.  In this work, we will develop a detrended fluctuation analysis (DFA)-based bivariate regression model to investigate this dependence.

The simplest and maturest method to describe the dependence of variables is the linear regression. However, the information gained from the traditional linear regression cannot fully meet our need of investigation on the dependence among different variables at different time periods. On the other hand, note that the DFA proposed in 1990s~\cite{PB1,PB2} performs excellently in analyzing the long-range correlations~\cite{KK} of a nonstationary series with fractality and multifractality~\cite{KJ,KZ} at different time-scales. To obtain the cross-correlation between two nonstationary series,   DFA was extended to the detrended cross-correlation analysis (DCCA)~\cite{PS}. By defining scale-dependent detrended fluctuation functions, the methods of DFA and DCCA together with their extensions have been applied in a wide range of disciplines~\cite{PH,WZ,WL1,WL2,WF1,WF2, WY, OD1,JZ,LS,KL1,KL2,YL,LY}. Since  the ordinary least squares (OLS) method expresses the estimated parameters of standard regression framework   as a form of variances and covariances,  it builds a bridge between the regression framework and the family of DFA/DCCA as the latter can also produce variances and covariances. Then, the idea of estimating multiple time scale regression coefficients can be achieved by the DFA/DCCA. Recently, Kristoufek~\cite{KL3} constructed a simple DFA-based regression framework exactly by this bridge. The selected examples show the relationship between the pair of variables varies strongly across scales.

In this work, we focus on the interaction of PM2.5 series of three adjacent cities in Northern China, namely, Beijing, Tianjin, and Baoding. The three cities form a triangle shape in the map. The distances between Beijing and Tianjin, Beijing and Baoding, and Tianjin and Baoding are about 115km, 140km, and 150km, respectively. All three cities have a population of more than 10 million  and have been greatly affected by heavy smog in recent years. The real-time data of PM 2.5 series of these three cities from December 1, 2013 to November 30, 2016 are chosen for our consideration, which are taken from the Ministry of Environmental Protection of the People's Republic of China (http://datacenter.mep.gov.cn). The original data show  an obvious periodic characteristic and roughly similar trends among the three cities, which imply that there is a possible relevance between per two cities of them. To verify that, the partial correlation technique is employed  to get the intrinsic relations between two cities by deleting the interference from the third variable. Four seasons, classified as winter (December, January, and February), spring (March, April, and May), summer (June, July, and August), and fall (September, October, and November), are considered.  The results are listed in Table~\ref{tab:1}.

\begin{table}[ht!]
\caption{Partial correlation coefficients and $t$-statistics between per two cities of Beijing, Tianjin, and Baoding in four seasons.}
\label{tab:1}
\begin{center}      
\begin{tabular}{lllll}
\hline\noalign{\smallskip}
\quad & winter & spring & summer & fall  \\
\noalign{\smallskip}\hline\noalign{\smallskip}
Beijing vs. Tianjin   & 0.3048 & 0.3072 & 0.2625 & 0.1660 \\
$t$-statistics & $25.8011^*$ & $26.2714^*$ & $22.1366^*$ & $13.2672^*$ \\
Beijing vs. Baoding & 0.2745 & 0.4545 & 0.4815 & 0.4468 \\
$t$-statistics & $23.0124^*$ & $41.5216^*$ & $44.7079^*$ & $39.3711^*$ \\
Tianjin vs. Baoding & 0.5570 & 0.4517 & 0.3461 & 0.5992 \\
$t$-statistics & $54.0752^*$ & $41.1961^*$ & $30.0154^*$ & $58.9941^*$ \\
\noalign{\smallskip}\hline
\end{tabular}\\
Note: * indicates statistical significance with 0.01 significance level.
\end{center}\end{table}

In Table~\ref{tab:1}, we also list the $t$-statistics ($t=r_{12,3}\sqrt{\frac{N-3}{1-r_{12,3}^2}}$, where $r_{12,3}$ denotes the partial correlation coefficient between the first and second variables eliminating the effects of the third one, $N-3$ is the degree of freedom) of the partial correlation coefficients to access the statistical significance at the given significance level. Unsurprisingly, Table~\ref{tab:1}  shows that the correlations of PM2.5 between both per two cities are of statistical significance. It explains that the air quality in one city of Northern China cannot be irrelevant to that of its neighbouring cities, which implies   potential dependence among the three cities. However, we also note in Table~\ref{tab:1} that the degree of relevance is different among different cities and in different seasons though all of them are significant.

To fully detect and quantify the dependence among the PM2.5 series of the above-mentioned  three cities, in this work, we construct a new bivariate regression framework which prevails the DFA method and allows us to investigate the dependence of three nonstationary series with multiple time scales. With the DFA-based variance instead of the standard variance, this new DFA bivariate regression model provides more information on the dependence among variables at different time scales. We organize the rest of this paper as follows. The performance of the proposed DFA regression model and the results on the application to PM2.5 series analysis are reported and discussed in the following section, which is followed by our conclusions. The methodologies including the standard regression method, the DFA method, and the DFA-based regression method are introduced at the end of this paper.

\section{Results and Discussions}
 \subsection*{Performance of DFA estimators}
The bivariate DFA-based regression model produces two time scale-based regression coefficients. This allows us to detect the dependence of a response variable and two independent variables at different time scales. In order to examine the validity of the model and show its advantages, in this section, we perform two numerical tests on the non-stationary bivariate regression frameworks $Y=\beta_0+\beta_1X_1+\beta_2X_2+\epsilon.$

In the first test, we investigate the performance of the DFA estimators under different levels of long-term dependence in $X_1$, $X_2$, and $Y$. According to ~\cite{KL3}, the setting I is given as below: two artificial series $X_1$ and $X_2$ with length $10000$ are generated by ARFIMA($0,d,0$) process with identical fractional integration parameter ($d$) and independent Gaussian noises ($\xi_i(t), i =1$ and 2 ) as $ X_i(t)=\sum_{n=0}^{\infty}a_n(d)\xi_i(t-n) $. The quantity $a_n(d)$ is defined by $a_n(d)=\Gamma(n-d)/[\Gamma(-d)\Gamma(n+1)]$, where $\Gamma(\cdot)$ is the Gamma function. The error-term $\epsilon$ is set as a standard Gaussian noise so that the response variable $Y$ has the same parameter $d$ as the two independent variables. The regression coefficients are set as $\beta_0 = \beta_1 = 1$ and $\beta_2 = 2$.  Fig.~1 shows mean values and standard deviation of the two DFA estimators $\beta_i^{DFA}$ ($i = $1 and 2) for the generated series with $d$ ranging from $-0.5$ to $0.5$ (at the step size of $0.1$). The estimators are averaged over scales between $10$ and $1000$ with a logarithmic isometric step. Each case is run $1000$ times to eliminate the noise interference. It is clear that the two estimators locate the two given regression coefficients of $1$ (Fig.~1a) and 2 (Fig.~1b) unbiasedly, and are independent of the value of $d$. In addition, the standard deviations of both estimators decrease with the increasing memory. The good performance shows that the method is feasible. On the other hand, to investigate the performance of the DFA estimators faced with a long-range dependent error-term $\epsilon$, we use  setting II   given as: the memory parameter $d$ is fixed at 0.4 for both   $X_1$ and $X_2$, and the $\epsilon$ is produced by an ARFIMA process with $d_\epsilon$ varying from $-0.5$ to $0.5$. Other settings are as those in setting I. Fig.~2 records similar information as that in Fig.~1. Although the fluctuation of DFA estimators increases with $d_\epsilon$, which is expected due to an increasing weight of the error-term in the dynamics of $Y$ with the increasing memory of the error-term, we are satisfied to find that the two estimators are still unbiased pointing to the given values with a  narrow range for each level of memory of the error-terms.

Our second numerical test aims to show that the DFA estimators are able to identify the dependence of studied variables at different time scales whereas  the classical method cannot. To this end, a binomial multifractal series (BMFs) is employed to be regarded as the independent variable $X_1$, which is constructed as  $X_1=p^{n-n[k-1]}(1-p)^{n[k-1]}$, $k=1$, $2$, $\ldots$,  $2^n$, where the parameter $p \in (0,0.5)$ (We take $p=0.3$ in our test), $n[k]$ denotes the number of digit $1$ in the binary representation of the index $k$. The variable $X_2$ is a  Gauss variable with $0$ mean and $0.0001$ standard deviation. Both $X_1$ and $X_2$ are of length $2^{15}$. The  bivariate regression framework $Y = \beta_0+\beta_1X_1+\beta_2X_2+\epsilon$ is set with the same coefficients as the first test ($\beta_0 = \beta_1 = 1$ and $\beta_2 = 2$). The error-term $\epsilon$ is the Gauss noise of the same strength as $X_2$. For the BMFs $X_1$, we remove all values smaller than $0.00001$ so that only a few of the largest elements are left. In their places, we substitute Gaussian distributed random numbers with $0$ mean and $0.0001$ standard deviation. Then we obtain a binomial cascade series embedded in random noise. We analyze the dependence between the response variable $Y$ and two independent variables and find that the estimated $\beta_2^{DFA}$ is unbiased at $2$ with a desirable error bar for every time scale, as shown in Fig. 3. However, the performance of $\beta_1^{DFA}$ has changed a lot. The dependence  between   $Y$ and $X_1$ is obviously less than the given value at the smaller scales contrary to the larger ones. This is because in the smaller scales, the dependency has been destroyed by the random noise. Our DFA estimators have the capability to recognize this effect while the classical estimators fail to do so (see the errorbar with circle symbol in Fig.~3).

\subsection*{Performance of the three models' regression coefficients}
As mentioned above, air pollution in Northern China is very serious in recent years. Fine particulate matter from industrial exhaust and smoke dust  forms smog to  fill  in the air. We now apply our DFA regression model to investigate the dependence of  PM2.5 series in these three cities. We build three bivariate models for Beijing, Tianjin, and Baoding, respectively. In Model I, the dependent variable ($Y$) is the PM2.5 of Beijing while the two independent variables are the PM2.5 of Tianjin ($X_1$) and Baoding ($X_2$); in Model II, $Y$ is the PM2.5 of Tianjin, $X_1$ is the PM2.5 series of Beijing and $X_2$ is the PM2.5 series of Baoding;    in Model III, $Y$ is the PM2.5 of Baoding,  $X_1$ and $X_2$ stand  for the PM2.5 series in Beijing and Tianjin, respectively. In this section, we first show the performance of the regression coefficients at different scales in the three models and then make two statistical tests for the two regression coefficients in each model. Some evaluations for  the DFA-based regression and the standard regression are conducted at the end of this section.

The two regression coefficient  estimators together with their standard deviations of the three models are sketched in Figs.~4--6, respectively. As expected, the effect is  obviously positive. However, a strong variation across scales is found in different seasons. More specifically,
 \begin{itemize}
\item[{(a)}] In the Beijing's model, Tianjin ($X_1$) has strongly positive effect  in every season,  especially for the larger time scales. On the contrary, Baoding ($X_2$) has different effects on Beijing. Compared to spring and summer, the effect is quite weak in the other two seasons, especially in winter,   $\beta_2^{DFA}(n)$ is nearly $0$ when the scale is more than $800$ hours.
\item[{(b)}] In the Tianjin's model, Baoding ($X_2$) presents more unstable effect at different scales. Particularly in summer,   $\beta_2^{DFA}(n)$ is close to $0$ from the smaller scale to the larger scale at about $50$ days ($1200$ hours), which implies that the positive correlation between Tianjin and Baoding can last less than $50$ days. In addition, the two coefficients are less than $0.5$ in most days, which indicates that Beijing and Baoding have little impact on the PM2.5 in Tianjin.
\item[{(c)}] For the model of Baoding, the  effect of Tianjin ($X_2$) to Baoding is similar to that of Baoding to Tianjin in model II. However, the fact that after approximately $17$ days ($408$ hours)  the effect reaches the value  greater than $1$ indicates  that an increase of $1$ unit PM2.5 concentration of Tianjin will lead to the increase of  more than $1$ unit PM2.5 concentration  in Baoding. In this regard, Tianjin has  more impact on Baoding. In addition, the narrow confidence intervals and low standard deviations (less than $0.02$) shown in all  sub-plots suggest satisfied reliability of the estimates.
\end{itemize}

\subsection*{Statistic significance tests of regression coefficients}
As mentioned above, the estimated $\hat{\beta}^{DFA}(n)$ is able to theoretically describe the dependence between the impulse variables and the response variables at different time scales. In theory, as long as   $\hat{\beta}^{DFA}_j(n)$  is not equal to zero, the independent variable $X_j$ will affect $Y$. However, for finite time series,  $\hat{\beta}^{DFA}_j(n)$  is not always equal to $0$ even in the absence of relationship between  $X_j$ and $Y$ due to the size limitation. Therefore, we perform a hypothesis test for the estimated $\hat{\beta}^{DFA}(n)$  to ensure the significance. The standard regression analysis provides a so-called $t$ statistic defined as $t_j=\frac{\hat{\beta}_j-\beta_j}{\sqrt{var(\hat{\beta}_j)}}$ ($j = 1$, 2)  for this purpose. We have $t_j\sim t(N-3)$ for the bivariate regression model as    $\hat{\beta}_j\sim N(\beta_j, var(\hat{\beta}_j))$. In general, if $\mid t_j\mid > t_{1-\alpha/2}(N-3)$ with a given $\alpha$, we should reject the null hypothesis of $\beta_j=0$ and the dependence between  $X_j$ and $Y$ is considered to be statistically significant. However, since lots of time scales are taken accounted in the DFA regression model, using a single critical value of $t_{1-\alpha/2}(N-3)$ is inappropriate. A correct way is to generate a critical value $t^c(n)$ for each time scale. To this end, inspired by the idea proposed by Podobnik et al.~\cite{PJ}, we shuffle the considered PM2.5 series and repeat the DFA regression calculations for $10,000$ times. Then let the integral of probability distribution function (PDF) from $-t^c(n)$ to $t^c(n)$ be equal to $1-\alpha$ (here, we take $\alpha = 0.01$). As an example, we show the PDF of $t^c(n)$ with five given $n$'s produced by the shuffled PM2.5 series of fall in Fig.~7.

As expected, the symmetrical PDF of $t^c(n)$ converges to a Gaussian distribution according to the central limit theorem. In addition, the critical value increases as $n$ increases. This implies that large time scale may strengthen dependence between two variables. By using  $t^c(n)$, we can determine whether the dependence between the impulse variable and the response variable is significant or not. In practice, the dependence between $X_j$ and $Y$ is present when $t_j(n)(=\frac{\hat{\beta}_j^{DFA}(n)-\beta_j}{\sqrt{var(\hat{\beta}_j^{DFA}(n))}})$ is larger than $t^c(n)$. For the four seasons, the scale-dependent $t$-statistics of regression coefficient together with the scale-dependent critical value $t^c(n)$ are presented in Fig.~8.

Note that in  Model I (for Beijing), the $t(n)$-statistics of $\beta_1^{DFA}(n)$ (Tianjin's coefficient) is equal to that of $\beta_1^{DFA}(n)$ (Beijing's coefficient) in   Model II (for Tianjin), the $t(n)$-statistics of $\beta_2^{DFA}(n)$ (Baoding's coefficient) is equal to that of $\beta_1^{DFA}(n)$ (Beijing's coefficient) in Model III (for Baoding), and in   Model II, the $t(n)$-statistics of Baoding's coefficient $\beta_2^{DFA}(n)$  is equal to that of Tianjin's coefficient $\beta_2^{DFA}(n)$ in Model III (for Baoding). Here the   three colored lines with different symbols represent the $t(n)$-statistics between each per two cities while the black dashed line stands for $t^c(n)$.
The partial DCCA coefficient $\rho_{PDCCA}(n)$ is recently developed
to uncover the intrinsic relation for two nonstationary series at different time scales. We also calculate the partial DCCA coefficients $\rho_{PDCCA}(n)$ of Beijing and Tianjin, Beijing and Baoding, and Tianjin and Baoding, respectively, and present the results  in Fig.~9. For the same purpose of testing the statistical significance, we also produce a critical value for the four seasons. Similarly, the PM2.5 data are shuffled $10,000$ times in the PDCCA calculations repeatedly, and thus $\rho_{PDCCA}^c(n)$ for 99\% confidence level is obtained, which is also shown in Fig.~9.

Comparing results in Fig.~8 and Fig.~9 gives  amazing similarities, which are also in agreement with the results shown in Figs.~4--6. Based on the results, we can draw the following three main points.
\begin{itemize}
\item[{(a)}] The dependence between Beijing and Tianjin (the blue square line)  gradually increases with the increasing time scales in all seasons. However, the dependence between the two cities is lower than other cities. This finding uncovers that  the reason for the serious air pollution in these two cities are  mainly due to their own heavy smog or are impacted by other cities.
\item[{(b)}] The dependence between Beijing and Baoding (the green triangle line) is significant in spring, summer, and fall. In winter, however, the dependence  disappears at long time scale, which implies that the two cities can only affect each other at  relatively short term. Moreover, compared to winter and fall, the dependence is much stronger in spring and summer, especially at long time scales, which indicates that they affect much longer in warm weather.
\item[{(c)}] In spring and summer, the $t(n)$-statistics and $\rho_{PDCCA}(n)$ of Tianjin vs. Baoding (the red circle line) go down through the critical lines of $t^c(n)$ and $\rho^c_{PDCCA}(n)$, respectively at about $800$ hours. This suggests that the dependence between Tianjin and Baoding will disappear when it's more than one month. However, the exact opposite occurs in winter and fall. In these two seasons, both   $t(n)$-statistic and $\rho_{PDCCA}(n)$ increase with the increasing time scales, which demonstrates that the interaction of bad air quality between the two cities will last longer in cold days.
\end{itemize}

\subsection*{Evaluations of DFA-based regression model} To evaluate our estimated DFA-based bivariate regression model, we plot the scale-dependent determination coefficient $R^2_{DFA}(n)$, and the beta coefficient $\beta^{*DFA}(n)$  and the average elasticity coefficient $\eta^{DFA}(n)$ in Fig.~10, and Figs.~11--13, respectively.

To show the new model provides  more information than the standard regression model does, we also include the three corresponding coefficients of standard bivariate regression model in these figures. As seen from  Fig.~10 that $R^2_{DFA}(n)$ is superior to the standard $R^2$ at most time scales. The good performance illustrates that one will gain richer information  in explaining the response variable when using our DFA-based regression model. On the other hand, we can conclude from Figs.~11--13 that (1) Baoding has more influence than Tianjin on Beijing in all seasons except for winter. (2) Tianjin is more sensitive to Baoding's changes in air quality than Beijing's in winter and fall. 
 (3) Tianjin affects Baoding more than Beijing  does in winter and fall, but less  in the other two seasons. In addition, Figs.~11--13 illustrate that the standard $\beta^*_j$ and $\eta_j$ can be seen as the mean values of the DFA-based $\beta^{*DFA}_j(n)$ and $\eta^{DFA}_j(n)$, respectively. This means that $\beta^{*DFA}_j(n)$ and $\eta^{DFA}_j(n)$ are able to measure the dependence degree of the studied independent variable on the dependent variable in all directions. Thus one can access the measurement according to his/her needs. For example, in winter of Model I, we find that the $\beta^{*DFA}_2(n)$ and $\eta^{DFA}_2(n)$ are larger than $\beta^{*DFA}_1(n)$ and $\eta^{DFA}_1(n)$, respectively, at smaller scales but much smaller at larger scales, which shows that the sensitivity of $Y$ to $X_2$ (Baoding) is greater than that of $Y$ to $X_1$ (Tianjin) for short term ($\leq 300$ hours) but Tianjin is more sensitive to   Beijing at the long term. This can help   air quality inspectors make the correct analysis for Beijing's PM2.5 at different periods.

\section{Conclusions}
The study of dependence between variables helps expose the causal relationship and correlation of the variables of interest in the real world. The linear regression model is undoubtedly one of the simplest methods among many approaches. However, single variety of regression coefficient and evaluation index cannot show all aspects of the dependence between independent variables and dependent variable. As a meaningful extension, we design a new framework for bivariate regression model using the prevailing DFA method. The proposed bivariate DFA regression model allows us to estimate multi-scale regression coefficients and other corresponding scale-dependent evaluation indicators. It has been shown via two artificial tests that these DFA-based regression coefficients are able to describe the dependence between the response variable and two independent variables exactly; and can capture different dependence at different time scales.

An application of the new model to the study of dependence of PM2.5 series among three heavily  air polluted cities in Northern China  unveils that huge difference of the dependence exists in per two cities in different seasons and at different periods. Three new indicators of the scale-dependent determination coefficient, the scale-dependent beta coefficient, and the scale-dependent elasticity coefficient are proposed, which turned out to be more practical than those in standard regression models. Three main points can be concluded as (1) Beijing and Baoding have little impact on the PM2.5 in Tianjin while Tianjin takes more impact on Baoding and the air quality of Beijing is more sensitive to the changes in Baoding. (2) In contrast, the air quality in Beijing and Tianjin is not significantly relevant, while the air quality in Tianjin and Baoding has a very significant impact on each other especially in the cold weather. (3) In comparison, the fluctuation of PM2.5 in Baoding has the greatest impact on the other two cities in most days. While Baoding's air quality is more sensitive to Beijing's changes in spring and summer, and is more sensitive to Tianjin's changes in winter and fall. These findings may provide some useful insights on understanding air pollution sources among cities in Northern China.

\section{Methods}

\subsection*{The standard bivariate regression model} To study the dependence of air quality among three neighboring cities, we consider a bivariate linear regression model as
\begin{equation}
Y=\beta_0+\beta_1X_1+\beta_2X_2+\epsilon,
\label{eq:Y}
\end{equation}
where $Y$ is a dependent variable, $X_1$ and $X_2$  are two independent variables, $\epsilon$ is a Gaussian error term with zero mean value, and $\beta_j$ ($j=1$, $2$) is the partial regression coefficient characterizing the dependence on $X_j$. The most critical work in empirical studies is to estimate $\beta_1$ and $\beta_2$.  The OLS method gives
\begin{equation}
\left\{
\begin{aligned}
\hat{\beta}_1 &=
\frac{(\sum^N_{t=1}x_{1t}y_t)\cdot(\sum_{t=1}^Nx_{2t}^2)-(\sum_{t=1}^Nx_{2t}y_t)\cdot(\sum_{t=1}^Nx_{1t}x_{2t})}
{(\sum_{t=1}^Nx_{1t}^2)\cdot(\sum_{t=1}^Nx_{2t}^2)-(\sum_{t=1}^Nx_{1t}x_{2t})^2} \sim \frac{\widehat{\sigma_{X_1Y}}\cdot\widehat{\sigma_{X_2}^2}
-\widehat{\sigma_{X_2Y}}\cdot\widehat{\sigma_{X_1X_2}}}
{\widehat{\sigma_{X_1}^2}\cdot\widehat{\sigma_{X_2}^2}-\widehat{\sigma_{X_1X_2}^2}^2},
\\
\hat{\beta}_2 &= \frac{(\sum^N_{t=1}x_{2t}y_t)\cdot(\sum_{t=1}^Nx_{1t}^2)-(\sum_{t=1}^Nx_{1t}y_t)\cdot(\sum_{t=1}^Nx_{1t}x_{2t})}
{(\sum_{t=1}^Nx_{1t}^2)\cdot(\sum_{t=1}^Nx_{2t}^2)-(\sum_{t=1}^Nx_{1t}x_{2t})^2} \sim \frac{\widehat{\sigma_{X_2Y}}\cdot\widehat{\sigma_{X_1}^2}-\widehat{\sigma_{X_1Y}}
\cdot\widehat{\sigma_{X_1X_2}}}{\widehat{\sigma_{X_1}^2}\cdot\widehat{\sigma_{X_2}^2}-\widehat{\sigma_{X_1X_2}^2}^2},
\end{aligned}
\right.
\label{eq:beta}
\end{equation}
where $\langle \cdot \rangle$ denotes the mean value of the whole time period, $x_{1t} = X_{1t}-\langle X_1\rangle$, $x_{2t} = X_{2t}-\langle X_2\rangle$, and $y_t = Y_t-\langle Y\rangle$. Then  the estimator of residuals can be determined by $\hat{e}_t=Y_t-\hat\beta_1X_{1t}-\hat\beta_2X_{2t}-\langle Y_t-\hat\beta_1X_{1t}-\hat\beta_2X_{2t}\rangle$. With it one can obtain the estimators of variance of the two regression coefficients as
\[
\left\{
\begin{aligned}
var(\hat\beta_1)  = \frac{(\sum_{t=1}^Nx_{2t}^2)\cdot\frac{\sum_{t=1}^N\hat{e}_t^2}{N-3}}
{(\sum_{t=1}^Nx_{1t}^2)\cdot(\sum_{t=1}^Nx_{2t}^2)-(\sum_{t=1}^Nx_{1t}x_{2t})^2} \sim  \frac{1}{N-3}\cdot \frac{\widehat{\sigma_{X_2}^2}\cdot\widehat{\sigma_\epsilon^2}}
{\widehat{\sigma_{X_1}^2}\cdot\widehat{\sigma_{X_2}^2}-\widehat{\sigma_{X_1X_2}^2}^2},
\\
var(\hat\beta_2) =  \frac{(\sum_{t=1}^Nx_{1t}^2)\cdot\frac{\sum_{t=1}^N\hat{e}_t^2}{N-3}}
{(\sum_{t=1}^Nx_{1t}^2)\cdot(\sum_{t=1}^Nx_{2t}^2)-(\sum_{t=1}^Nx_{1t}x_{2t})^2} \sim \frac{1}{N-3}\cdot \frac{\widehat{\sigma_{X_1}^2}\cdot\widehat{\sigma_\epsilon^2}}
{\widehat{\sigma_{X_1}^2}\cdot\widehat{\sigma_{X_2}^2}-\widehat{\sigma_{X_1X_2}^2}^2}.
\end{aligned}
\right.
\]

The variance illustrates the accuracy of the estimated parameters. The estimated regression coefficients together with their variances can be further employed for hypothesis test and model evaluation. As an important indicator to evaluate the regression model, the determination coefficient $R^2$ is defined by
\begin{equation}
R^2=1-\frac{\sum^N_{t=1}\widehat{e_t}^2}{\sum^N_{t=1}y_t^2}=1-\frac{\widehat{\sigma_\epsilon^2}}{\widehat{\sigma_Y^2}},
\label{eq:R}
\end{equation}
with the range of $[0,1]$. $R^2$ measures a proportion of variance of $Y$ explained by $X_1$ and $X_2$ and higher value of $R^2$ implies better model interpretation ability. Moreover, to quantify sensitivity of explained variable to each explaining variable, two quantities, namely, the beta coefficient (denoted as $\beta_j^*$) and the average elasticity coefficient (denoted as $\eta_j$), are defined
\begin{equation}
\beta_j^*=\hat{\beta_j}\sqrt{\frac{\sum^N_{t=1}x_{jt}^2}{\sum^N_{t=1}y_t^2}} \qquad \mbox{for $j =1$  and $2$}
\label{eq:beta*}
\end{equation}
and
\begin{equation}
\eta_j=\hat{\beta_j}\frac{\langle X_j\rangle}{\langle Y\rangle} \qquad \mbox{for $j =1$  and $2$,}
\label{eq:eta}
\end{equation}
which can explain the relative importance of variables $X_1$ and $X_2$ to $Y$. According to~\cite{KL3}, the advantage of translating the standard notation into variance and covariance shown on the right-hand side of Eqs.~(\ref{eq:beta})-(\ref{eq:eta}) is available to use the DFA/DCCA methods based on the same idea.

\subsection*{The DFA-based variance and DCCA-based covariance functions} DFA and DCCA methods are described as follows. For a time series $\{z_t\}$, $t=1$, 2, $\cdots$, $N$, we split its profile $Z_t=\sum_{i=1}^t(z_i-\langle z\rangle)$ into $N_n=[N/n]$ nonoverlapping segments of equal length $n$, denoted as $Z_{j,k}$, $k=1$, 2, $\cdots$, $n$. The same procedure is repeated starting from the opposite end to avoid disregarding a short part of the series in the end and thus $2N_n$ segments are obtained altogether. In the $j^{th}$ segment, we have $Z_{j,k}=Z_{(j-1)n+k}$ for $j=1$, 2, $\cdots$, $N_n$ and $Z_{j,k}=Z_{N-(j-N_n)n+k}$ for $j=N_n+1$, $N_n+2$, $\cdots$, $2N_n$, where $k=1$, 2, $\cdots$, $n$. In each segment, the local linear (or other) trend ~\cite{OD2, LB} can be fitted as $\widehat{X_{j,k}}$ (in our work, we use $2^{nd}$ order polynomial to fit the trend in each segment). Fluctuation function $f_Z^2(n,j)$ is then defined for each segment as
\begin{equation}
f_Z^2(n,j)=\frac{1}{n}\sum_{k=1}^n(Z_{j,k}-\widehat{Z_{j,k}})^2.
\label{eq:fZ}
\end{equation}
Averaging the fluctuation $f_Z^2(n,j)$ over all segments yields
\begin{equation}
F_Z^2(n)=\frac{1}{2N_n}\sum_{j=1}^{2N_n}f_Z^2(n,j),
\label{eq:FZ}
\end{equation}
which is the so-called DFA-based scale-dependent variance function. To obtain the scale-dependent covariance of two equal length series $\{z_{1t}\}$ and $\{z_{2t}\}$, $t=1$, 2, $\cdots$, $N$, we only need to translate the univariate fluctuation function in each segment and average fluctuation into the bivariate case directly,
\begin{equation}
f_{Z_1Z_2}^2(n,j)  =  \frac{1}{n}\sum_{k=1}^n(Z_{1j,k}-\widehat{Z_{1j,k}})(Z_{2j,k}-\widehat{Z_{2j,k}}),
\label{eq:fZ1Z2}
\end{equation}
\begin{equation}
F_{Z_1Z_2}^2(n) =  \frac{1}{2N_n}\sum_{j=1}^{2N_n}f_{Z_1Z_2}^2(n,j).
\label{eq:FZ1Z2}
\end{equation}
The scale-characteristic fluctuation $F_{Z_1Z_2}^2(n)$ is the so-called DCCA-based covariance, which expresses the cross-correlation fluctuations between the series of $\{z_{1t}\}$ and $\{z_{2t}\}$. Thus we have obtained all objects to create the DFA-based regression model. But for purpose of testing, we need some accessories of the DFA process. The DCCA cross-correlation coefficient $\rho(n)$, proposed by Zebende~\cite{GZ}, can measure the cross-correlation between two nonstationary series at multiple time scales, which is defined as
\begin{equation}
\rho_{DCCA}(Z_1,Z_2,n)=\frac{F_{Z_1Z_2}^2(n)}{\sqrt{F_{Z_1}^2(n)F_{Z_2}^2(n)}}.
\label{eq:rho}
\end{equation}
To access intrinsic relations between the two time series on time scales of $n$, Yuan et al.~\cite{YF} and Qian et al.~\cite{QL} developed a so-called partial DCCA coefficient independently, which applies partial correlation technique to delete the impact of other variables on the two currently studied variables. This coefficient is  defined as
\begin{equation}
\rho_{PDCCA}(Z_1,Z_2,n)=-\frac{C_{j_1,j_2}(n)}{\sqrt{C_{j_1,j_1}(n)C_{j_2,j_2}(n)}},
\label{eq:p_rho}
\end{equation}
where $C$ is the inverse matrix of the cross-correlation matrix produced by $\rho_{DCCA}(n)$ of $Z_1$, $Z_2$, $\cdots$, and subscripts $j_1$ and $j_2$ stand respectively for the row and column of the location of $\rho_{DCCA}(Z_1,Z_2,n)$.

\subsection*{The DFA-based bivariate regression model} We now translate the standard bivariate regression process described above into the DFA-based bivariate regression model. The two estimators in Eq.~(\ref{eq:beta}) can be extended to  the scale-dependent estimators in the following way using the scale-dependent variance and covariance defined in Eqs.~(\ref{eq:FZ}) and~(\ref{eq:FZ1Z2}),
\begin{equation}
\left\{\begin{array}{rcl}
\hat{\beta_1}^{DFA}(n) =  \frac{F_{X_1Y}^2(n)\cdot F_{X_2}^2(n)-F_{X_2Y}^2(n)\cdot F_{X_1X_2}^2(n)}{F_{X_1}^2(n)\cdot F_{X_2}^2(n)-[F_{X_1X_2}^2(n)]^2},
\\
\hat{\beta_2}^{DFA}(n)  =  \frac{F_{X_2Y}^2(n)\cdot F_{X_1}^2(n)-F_{X_1Y}^2(n)\cdot F_{X_1X_2}^2(n)}{F_{X_1}^2(n)\cdot F_{X_2}^2(n)-[F_{X_1X_2}^2(n)]^2}.
\end{array}\right.
\label{eq:Vbeta}
\end{equation}
Similarly, the scale-dependent residuals are
\[
\hat{e_t}(n)=Y_t-\hat\beta_1^{DFA}(n)X_{1t}-\hat\beta_2^{DFA}(n)X_{2t}-\langle Y_t-\hat\beta_1^{DFA}(n)X_{1t}-\hat\beta_2^{DFA}(n)X_{2t}\rangle
\]
with zero mean value. Inserting the calculated $\hat{e_t}(n)$ into the DFA process, we obtain the fluctuation $F_\epsilon^2(n)$ to estimate the variances of $\hat{\beta_1}^{DFA}(n)$ and $\hat{\beta_2}^{DFA}(n)$ via Eq.~(\ref{eq:Vbeta}) as
\begin{equation}
\left\{\begin{array}{rcl}
var(\hat{\beta_1}^{DFA}(n))  =  \frac{1}{N-3}\cdot \frac{F_{X_2}^2(n)\cdot F_\epsilon^2(n)}
{F_{X_1}^2(n)\cdot F_{X_2}^2(n)-[F_{X_1X_2}^2(n)]^2},
\\
var(\hat{\beta_2}^{DFA}(n))  =  \frac{1}{N-3}\cdot \frac{F_{X_1}^2(n)\cdot F_\epsilon^2(n)}
{F_{X_1}^2(n)\cdot F_{X_2}^2(n)-[F_{X_1X_2}^2(n)]^2}.
\end{array}\right.
\label{eq:VbetaDFA}
\end{equation}
Then Eqs.~(\ref{eq:R})-(\ref{eq:eta}) can be translated into the DFA regression form as
\begin{equation}
R_{DFA}^2(n) =  1-\frac{F_\epsilon^2(n)}{F_Y^2(n)},
\label{eq:RDFA}
\end{equation}
\begin{equation}
\beta_j^{*DFA}(n)  =  \hat{\beta_j}^{DFA}(n)\sqrt{\frac{F_{X_j}^2(n)}{F_Y^2(n)}} \qquad \mbox{for $j=1$ and $2$,}
\label{eq:betaDFA}
\end{equation}
and
\begin{equation}
\eta_j^{DFA}=\hat{\beta_j}^{DFA}(n)\frac{\langle X_j\rangle}{\langle Y\rangle} \qquad \mbox{for $j=1$ and $2$.}
\label{eq:etaDFA}
\end{equation}

Comparing to the standard $R^2$, $\beta^*$, and $\eta$, the scale-dependent $R^2_{DFA}(n)$, $\beta^{*DFA}(n)$, and  $\eta^{DFA}(n)$ express more abundant information on model interpretation from multiple time scales.


\section*{Acknowledgements}
 This work was partially supported by National Natural Science Foundation of China (No.31501227, No. 11401577), and NSERC of Canada. The authors would like to thank two anonymous reviewers and the handling editor for their comments and suggestions, which led to a great improvement to the presentation of this work.

\section*{Author Contributions}

F.W. designed the framework and performed the statistical analysis. L.W. wrote the manuscript and checked the analysis. Y.C. reviewed the analysis.

\section*{Additional Information}
{\bf Competing Interests:} The authors declare no competing interests.

\begin{figure}
\includegraphics[angle=0, width=0.9\textwidth]{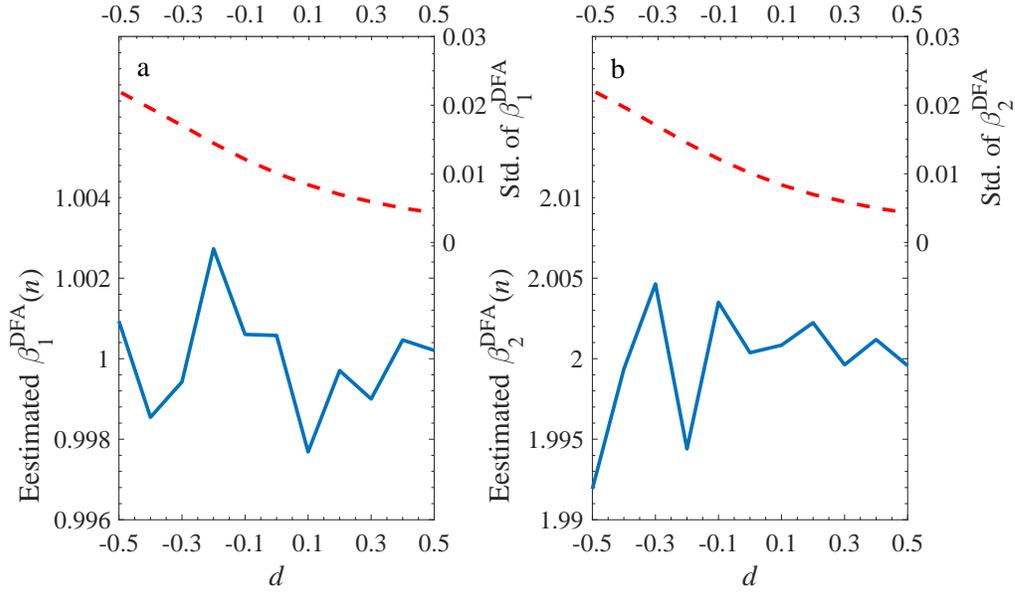}
\caption{Bivariate DFA regression of Beijing. Main planes of subplots (a), (b), (c), and (d) show estimated DFA regression coefficients $\beta_1(n)$ and $\beta_2(n)$ of winter, spring, summer, and fall, respectively. Gray zones denote  95\% confidence intervals. Inserts   are standard deviations of  $\hat{\beta}^{DFA}_1(n)$ and $\hat{\beta}^{DFA}_2(n)$. Subscripts 1 and 2 denote Tianjin and Baoding, respectively.}
\label{fig:BJ}       
\end{figure}

\begin{figure}
\includegraphics[angle=0, width=0.9\textwidth]{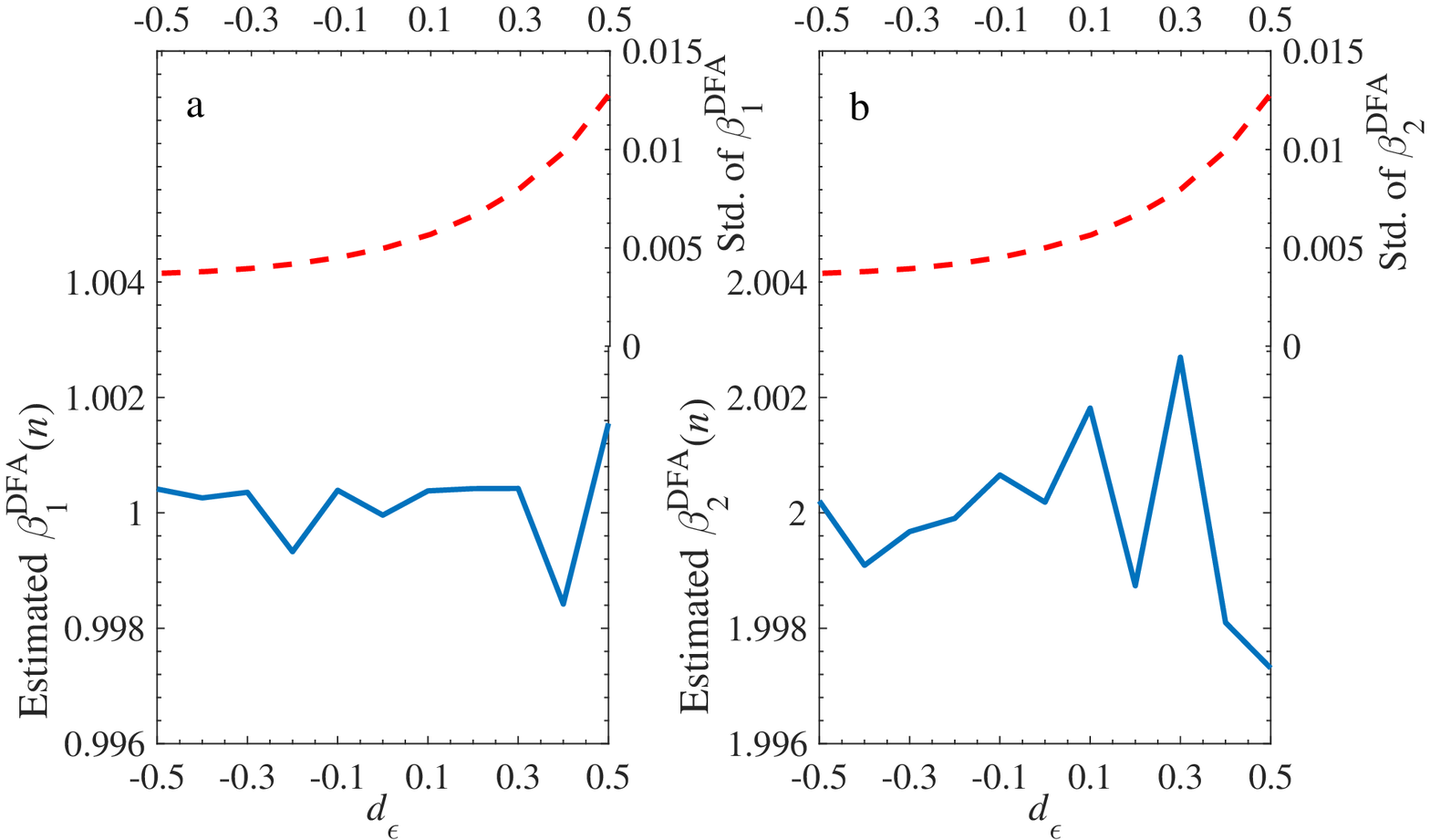}
\caption{Bivariate DFA regression of Tianjin with the same legend as in Fig.~\ref{fig:BJ}. Here, subscripts 1 and 2 denote Beijing and Baoding, respectively.}
\label{fig:TJ}       
\end{figure}

\begin{figure}
\includegraphics[angle=0, width=0.9\textwidth]{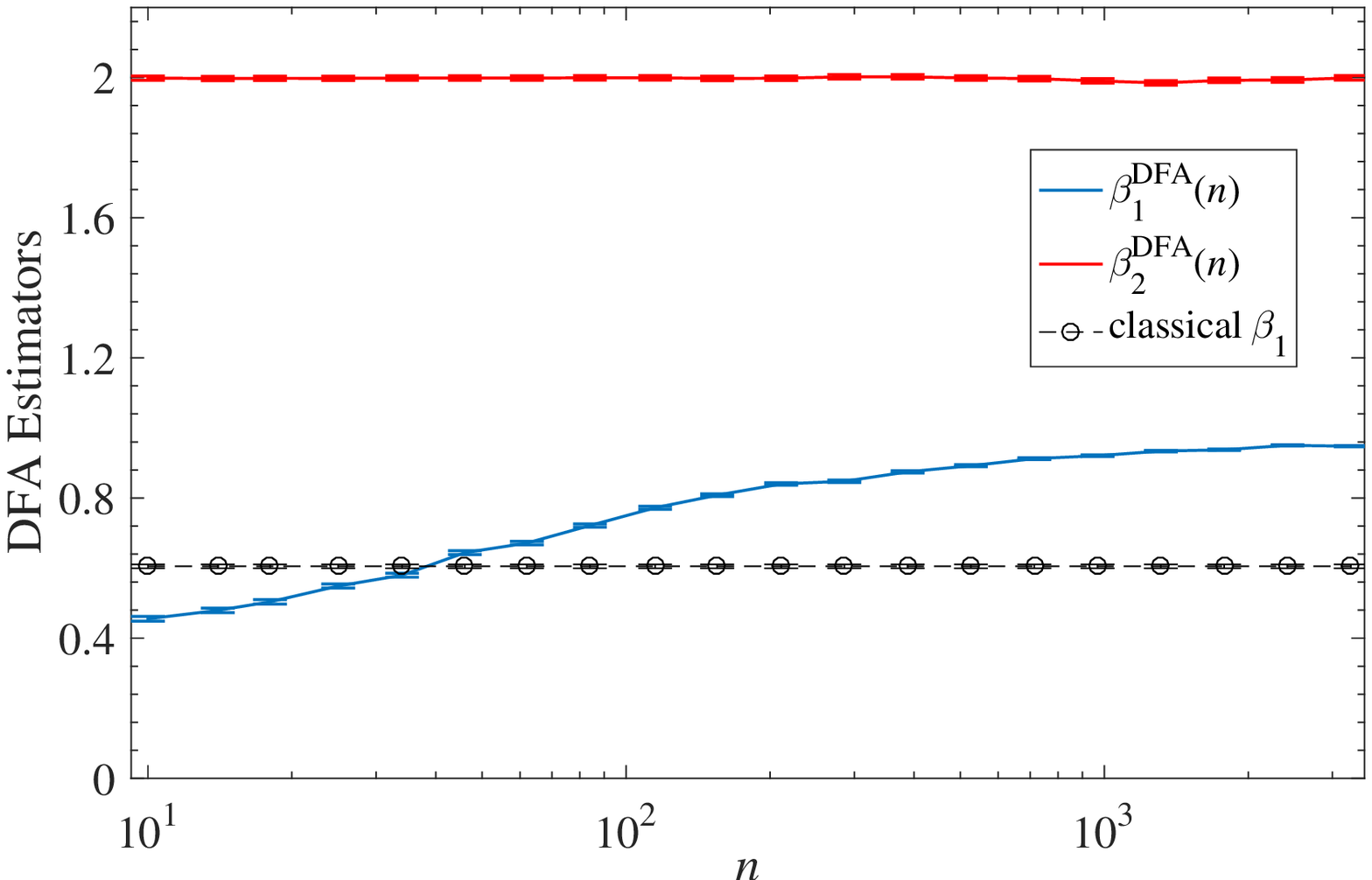}
\caption{Bivariate DFA regression of Baoding with the same legend as in Fig.~\ref{fig:BJ}. Here, subscripts 1 and 2 denote Beijing and Tianjin, respectively.}
\label{fig:BD}       
\end{figure}

\begin{figure}[ht!]
\includegraphics[angle=0, width=0.7 \textwidth]{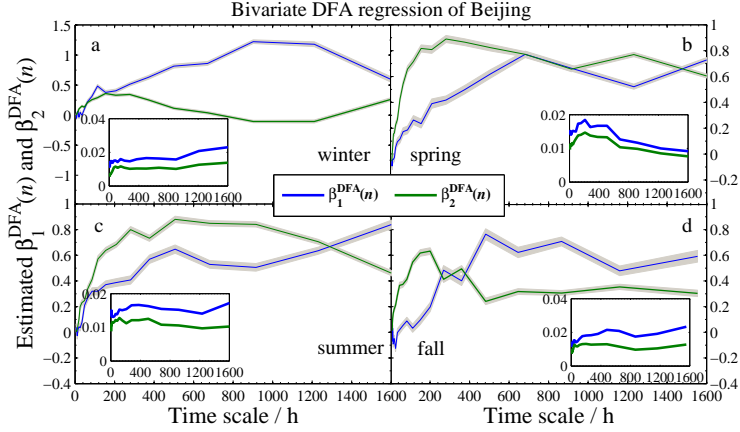}
\caption{PDF of critical points $t$- statistics critical values at different scales for the statistical test with $10000$ times of the shuffled PM $2.5$ series of fall.}
\label{fig:PDF}       
\end{figure}

\begin{figure}
\includegraphics[angle=0, width=0.9\textwidth]{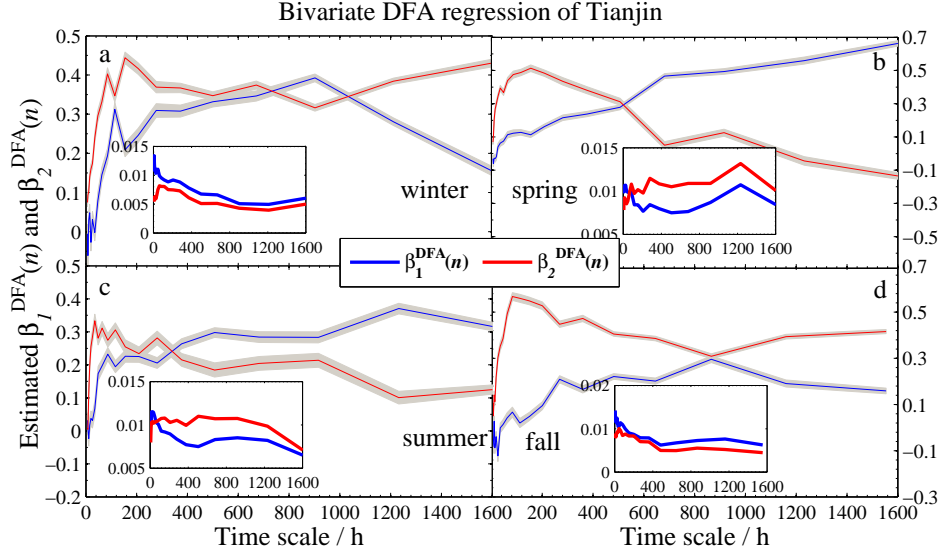}
\caption{$t$-statistical test of the estimated DFA-based bivariate regression coefficients. (a)-(d) are for winter, spring, summer and fall, respectively. The dashed line represents the $t^c(n)$ with $0.01$ significant levels. Above this line means decline of the null hypothesis  $\beta_j=0$.}
\label{fig:Tc}       
\end{figure}

\begin{figure}
\includegraphics[angle=0, width=0.9\textwidth]{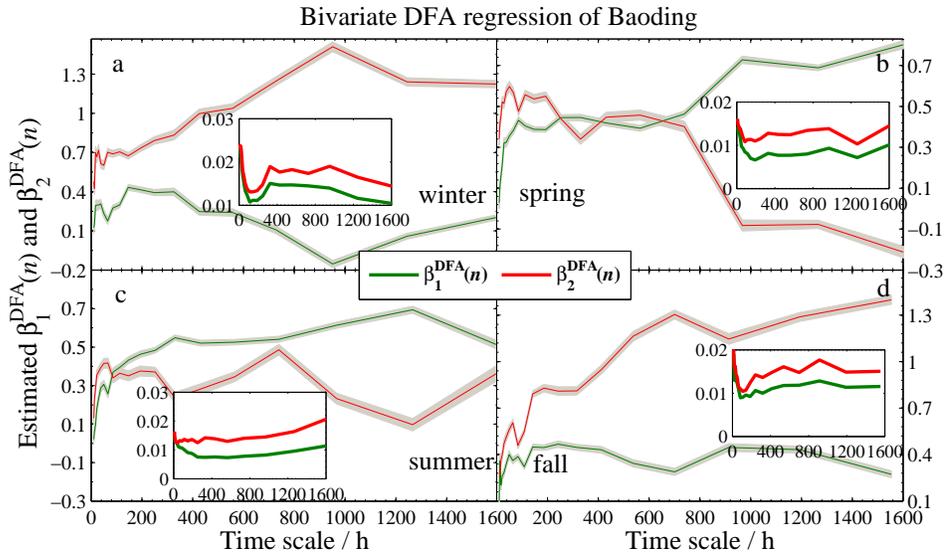}
\caption{Statistical test of DPCCA coefficients among the three cities. (a)-(d) are for winter, spring, summer, and fall, respectively. The dashed line represents the critical value  of $\rho_{DPCCA}$ which is obtained from $10000$ times Monte-Carlo simulations with 99\% confidence level. Below this line suggests no cross-correlated significance. }
\label{fig:PR}       
\end{figure}

\begin{figure}
\includegraphics[angle=0, width=1\textwidth]{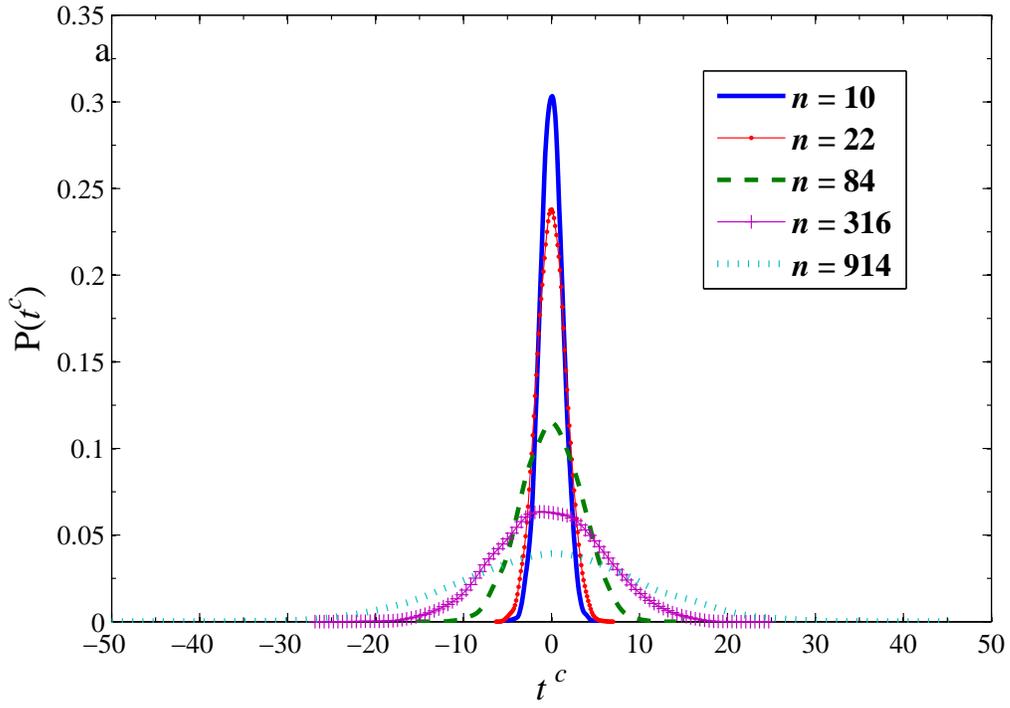}
\caption{Determination coefficients of bivariate DFA and standard regression model. (a)-(d) are for models of Beijing, (e)-(h) are for models of Tianjin, and (i)-(l) are for models of Baoding. The solid line denotes $R_{DFA}^2(n)$ and the dashed line denotes $R^2$.}
\label{fig:RRDFA}       
\end{figure}

\begin{figure}
\includegraphics[angle=0, width=1\textwidth]{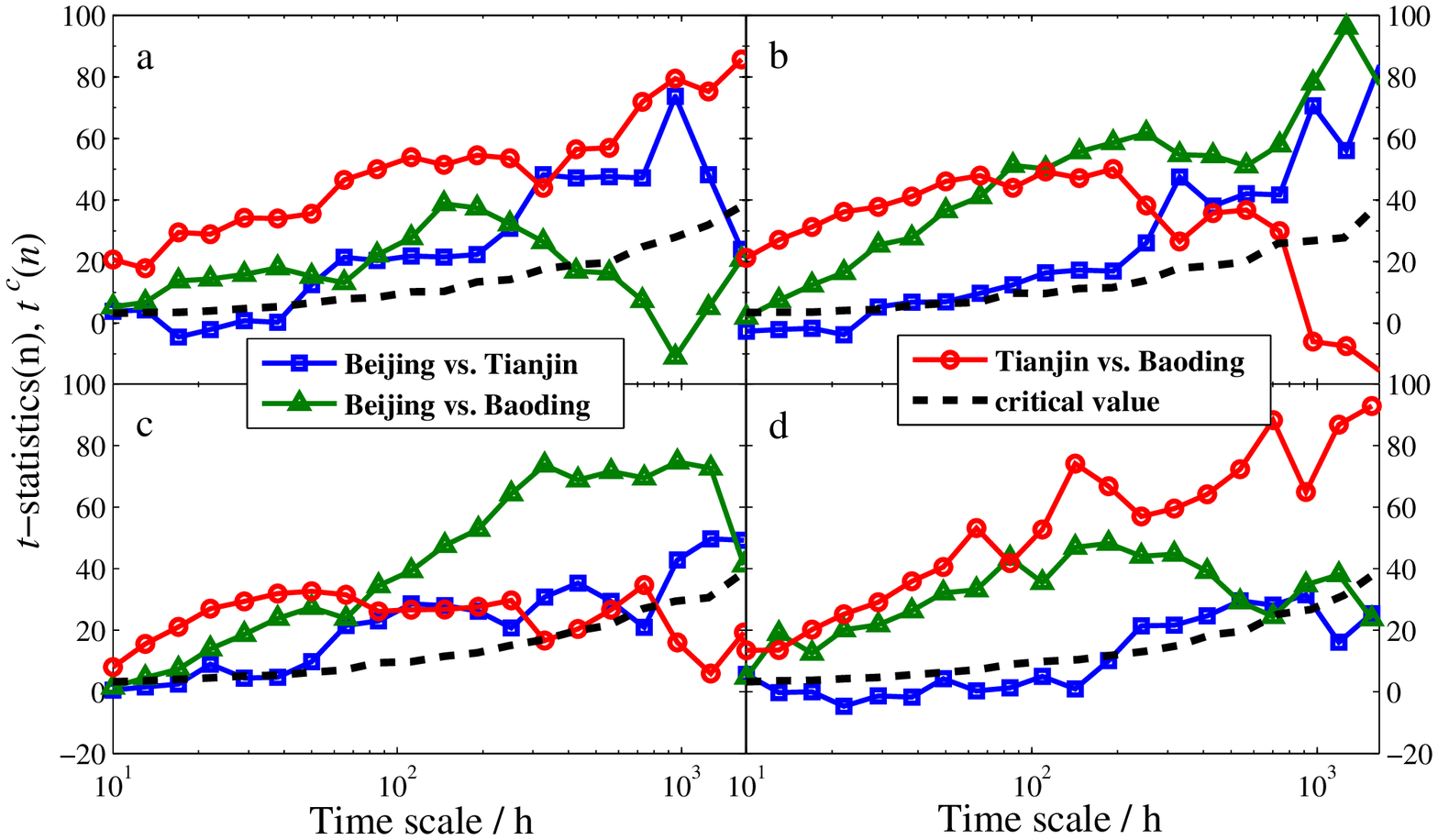}
\caption{Beta coefficients and elasticity coefficients of bivariate DFA and standard regression model of Beijing. The four columns from left to right are for winter, spring, summer, and fall, respectively. The subscript 1 of $\beta$ and $\eta$ denotes Tianjin and 2 denotes Baoding. }
\label{fig:BJ_be}       
\end{figure}

\begin{figure}
\includegraphics[angle=0, width=1\textwidth]{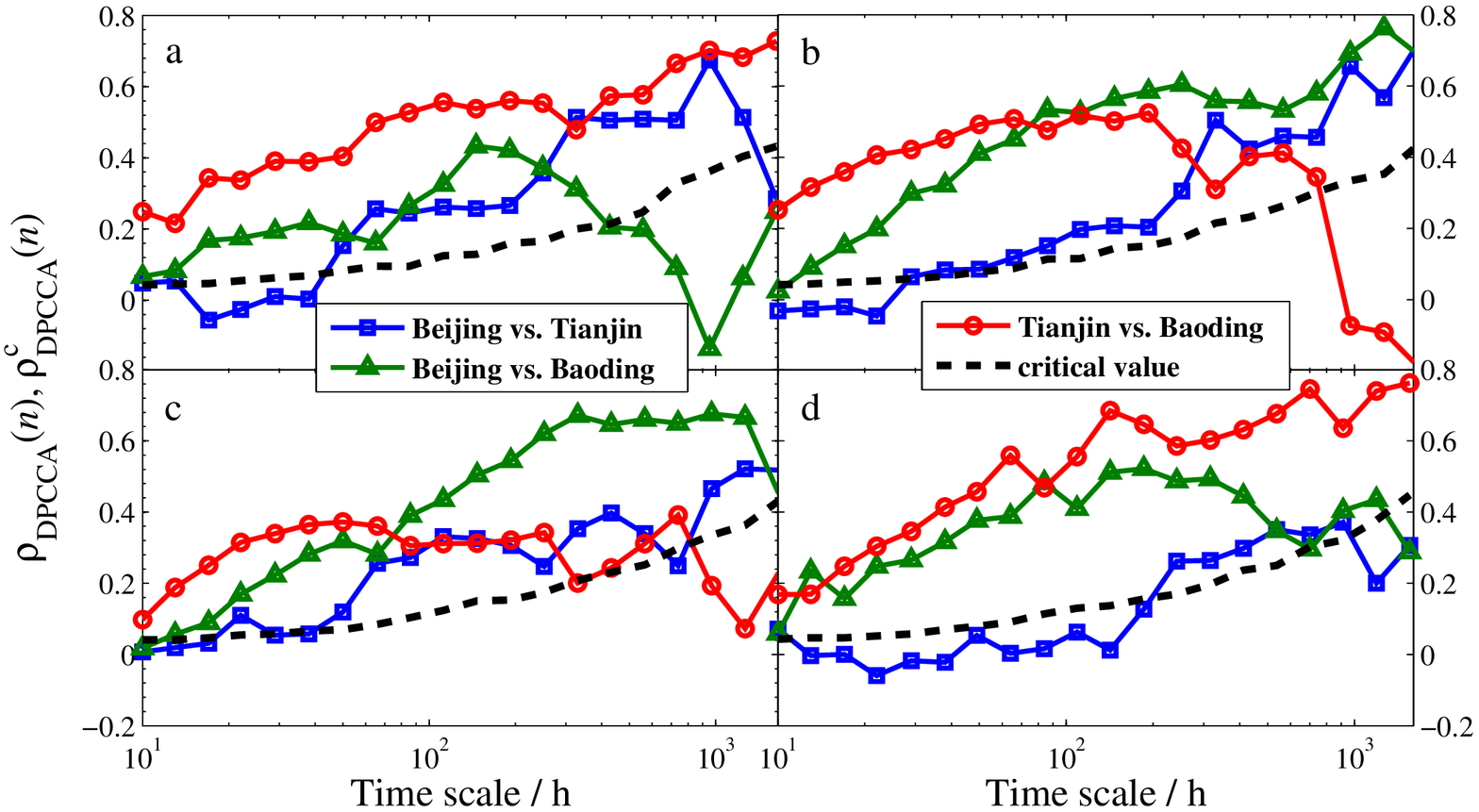}
\caption{Beta coefficients and elasticity coefficients of bivariate DFA and standard regression model of Tianjin with the same legend as in Fig.~\ref{fig:BJ_be}. Here the subscripts 1 and 2 denote Beijing and Baoding, respectively.}
\label{fig:TJ_be}       
\end{figure}

\begin{figure}
\includegraphics[angle=0, width=1\textwidth]{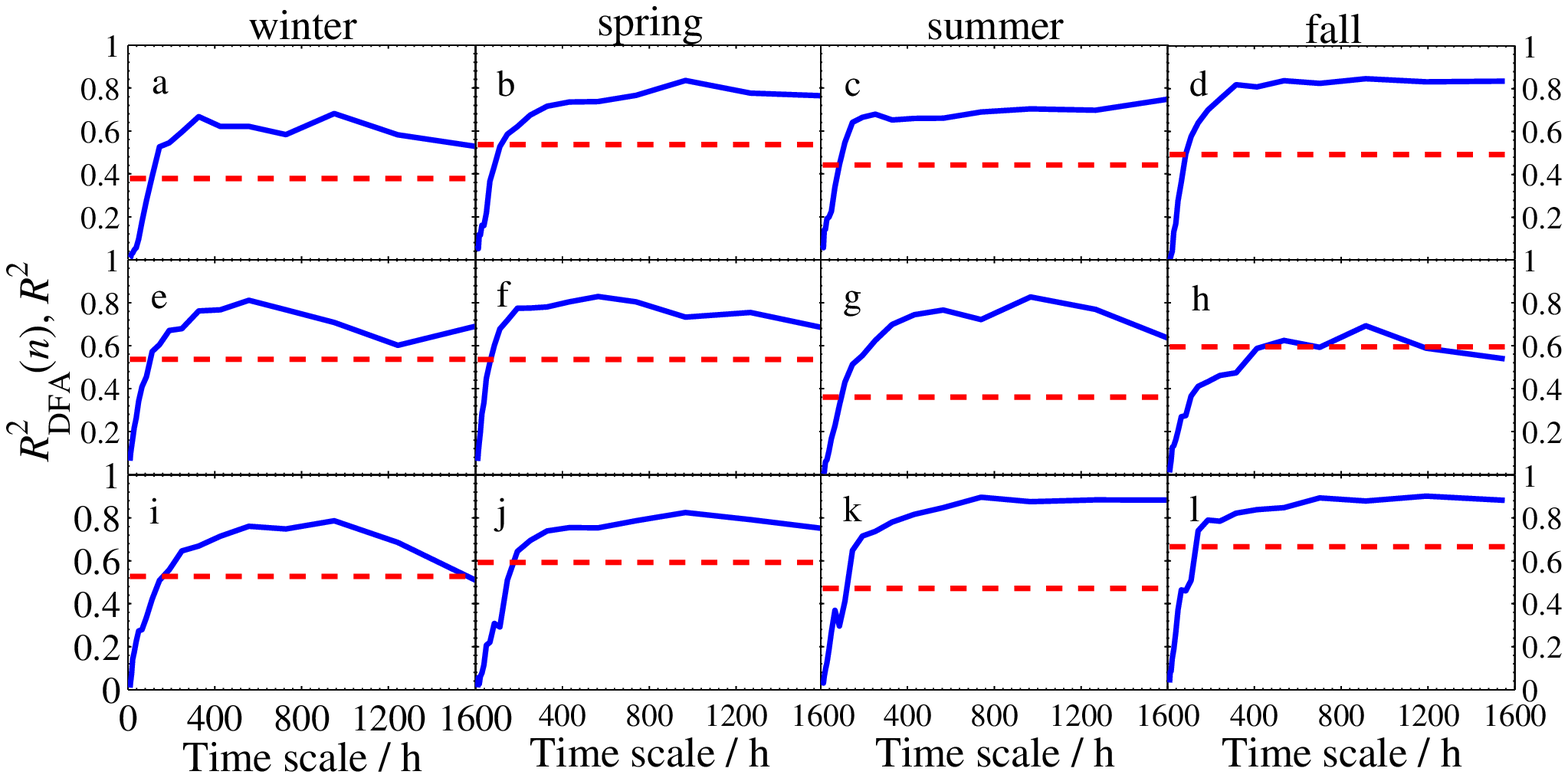}
\caption{Beta coefficients and elasticity coefficients of bivariate DFA and standard regression model of Baoding with the same legend as in Fig.~\ref{fig:BJ_be}. Here the subscripts 1 and 2 denote Beijing and Tianjin, respectively. }
\label{fig:BD_be}       
\end{figure}

\begin{figure}
\includegraphics[angle=0, width=1\textwidth]{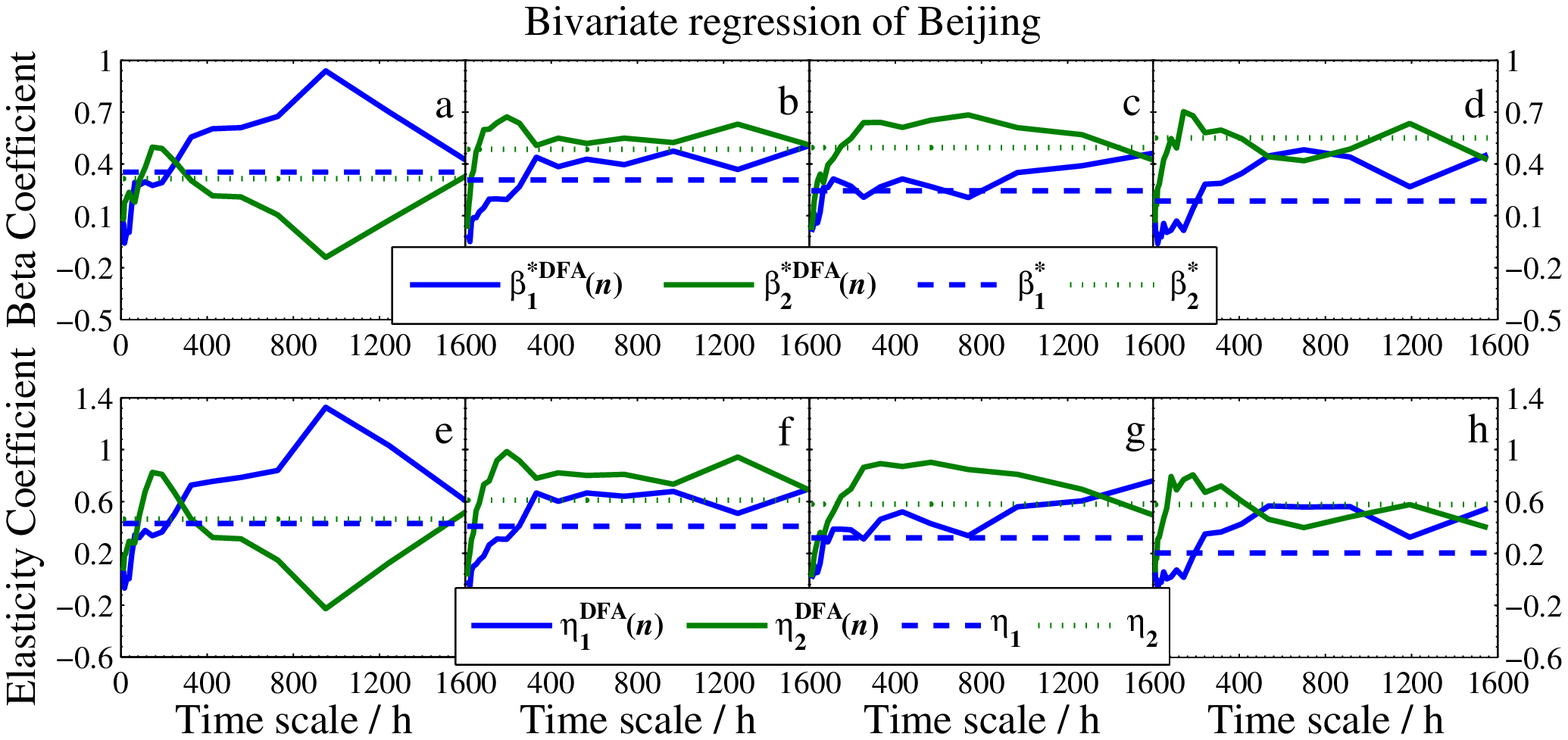}
\caption{Beta coefficients and elasticity coefficients of bivariate DFA and standard regression model of Beijing. The four columns from left to right are for winter, spring, summer, and fall, respectively. The subscript 1 of $\beta^*$ and $\eta$ denotes Tianjin and 2 denotes Baoding.}
\label{fig11}       
\end{figure}

\begin{figure}
\includegraphics[angle=0, width=1\textwidth]{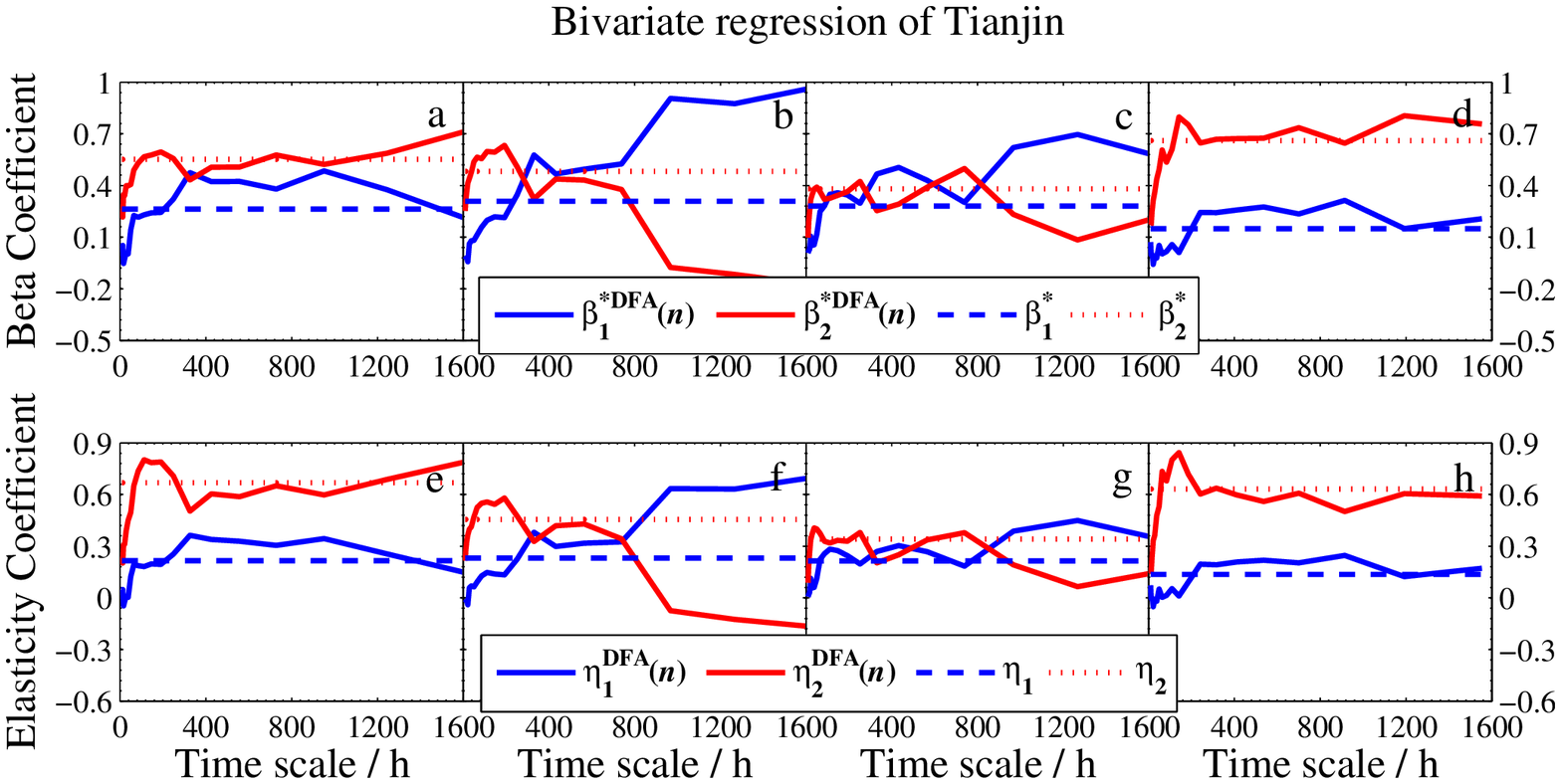}
\caption{Beta coefficients and elasticity coefficients of bivariate DFA and standard regression model of Tianjin with the same legend as in Fig.~11. Here the subscripts 1 and 2 denote Beijing and Baoding, respectively.}
\end{figure}

\begin{figure}
\includegraphics[angle=0, width=1\textwidth]{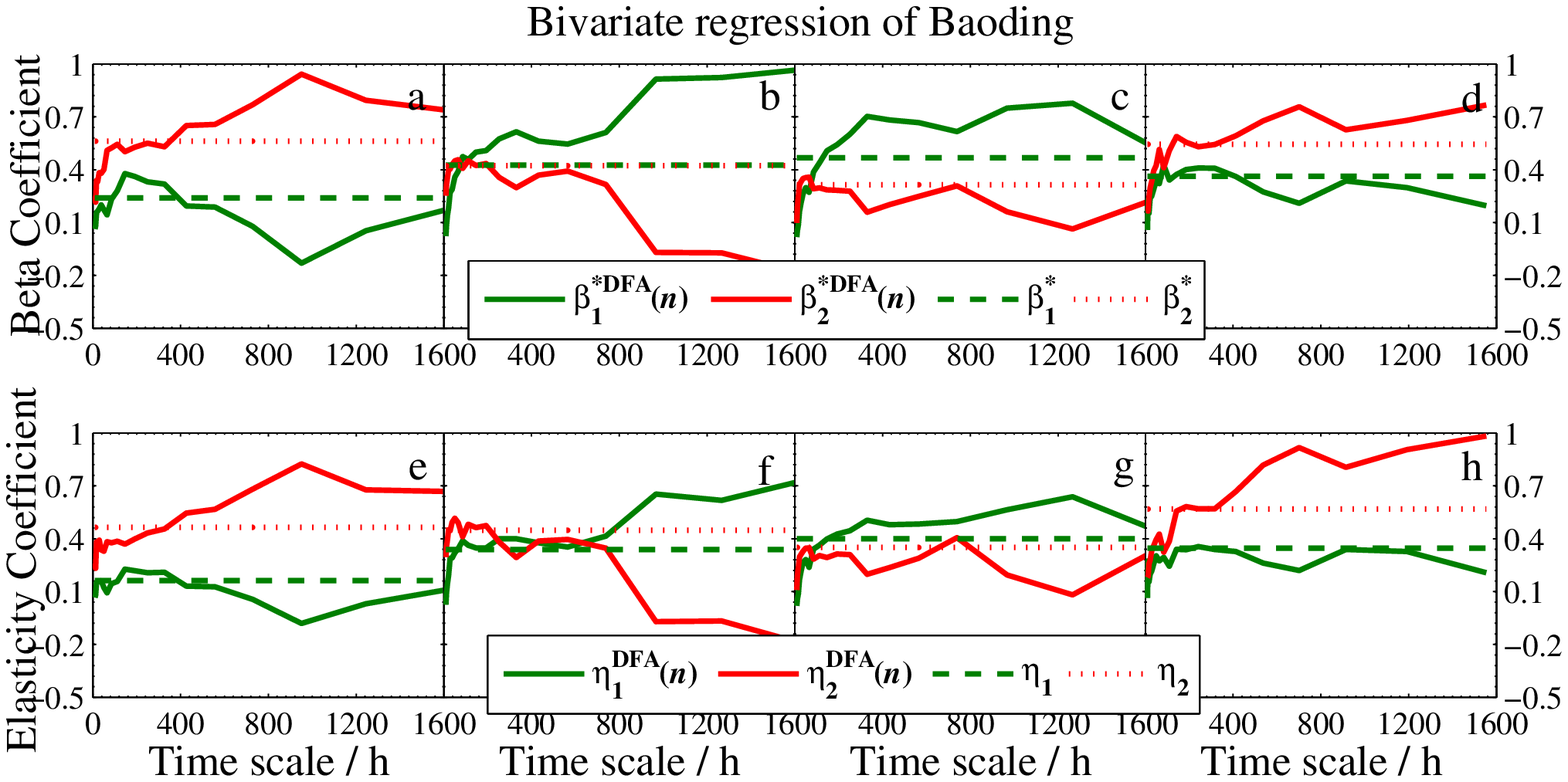}
\caption{Beta coefficients and elasticity coefficients of bivariate DFA and standard regression model of Baoding with the same legend as in Fig.~11. Here the subscripts 1 and 2 denote Beijing and Tianjin, respectively.}
\end{figure}

\end{document}